\documentclass[5p]{elsarticle}

\usepackage{lineno}

\usepackage{amssymb}
\usepackage{amsmath}
\usepackage[bookmarks=false,colorlinks=true,linkcolor=blue,citecolor=blue,filecolor=blue,urlcolor=blue]{hyperref}
\usepackage{tabularx}

\usepackage{dcolumn}
\usepackage{color}
\journal{Nuclear Instruments and Methods in Physics Research A}

\biboptions{sort&compress}

\begin{document}

\begin{frontmatter}

\title{Status of the JENSA gas-jet target for experiments with rare isotope beams}
\tnotetext[mytitlenote]{\copyright 2018. This manuscript version is made available under the \href{http://creativecommons.org/licenses/by-nc-nd/4.0/}{CC-BY-NC-ND 4.0 license}}

\author[1,2]{K.~Schmidt \corref{mycorrespondingauthor}}
\cortext[mycorrespondingauthor]{Corresponding author: Konrad Schmidt, National Superconducting Cyclotron Laboratory, Michigan State University, 640 South Shaw Lane, East Lansing, MI 48824, USA; \emph{present address}: TU Dresden, Zellescher Weg 19, 01069 Dresden; \emph{Email address:} \href{mailto:schmidtk@nscl.msu.edu}{Konrad.Schmidt1@tu-dresden.de}}
\author[3]{K.\,A.~Chipps}
\author[1,2]{S.~Ahn}
\author[4]{D.\,W.~Bardayan}
\author[1,5]{J.~Browne}
\author[6]{U.~Greife}
\author[7]{Z.~Meisel}
\author[1]{F.~Montes}
\author[4]{P.\,D.~O'Malley}
\author[1,5]{W-J.~Ong}
\author[3]{S.\,D.~Pain}
\author[1,2,5]{H.~Schatz}
\author[8]{K.~Smith}
\author[3]{M.\,S.~Smith}
\author[9]{P.\,J.~Thompson}
\address[1]{National Superconducting Cyclotron Laboratory, Michigan State University, East Lansing, USA}
\address[2]{JINA-CEE, Michigan State University, East Lansing, MI, USA}
\address[3]{Physics Division, Oak Ridge National Laboratory, Oak Ridge, TN, USA}
\address[4]{Institute for Structure and Nuclear Astrophysics, University of Notre Dame, Notre Dame, IN, USA}
\address[5]{Department of Physics and Astronomy, Michigan State University, East Lansing, MI, USA}
\address[6]{Department of Physics, Colorado School of Mines, Golden, CO, USA}
\address[7]{Department of Physics \& Astronomy, Ohio University, Athens, OH, USA}
\address[8]{Los Alamos National Laboratory, Los Alamos, NM, USA}
\address[9]{Department of Physics \& Astronomy, University of Tennessee, Knoxville, TN, USA}

\begin{abstract}
The JENSA gas-jet target was designed for experiments with radioactive beams provided by the rare isotope re-accelerator ReA3 at the National Superconducting Cyclotron Laboratory. The gas jet will be the main target for the Separator for Capture Reactions SECAR at the Facility for Rare Isotope Beams on the campus of Michigan State University, USA. In this work, we describe the advantages of a gas-jet target, detail the current recirculating gas system, and report recent measurements of helium jet thicknesses of up to about $10^{19}$ atoms/cm$^2$. Finally a comparison with other supersonic gas-jet targets is presented.
\end{abstract}


\end{frontmatter}


\section{Introduction\label{sec:introduction}}

The Jet Experiments in Nuclear Structure and Astrophysics (JENSA) windowless gas-jet target has been installed at the rare isotope beam facility ReA3 \cite{Leitner2013NIMB} at NSCL (National Superconducting Cyclotron Laboratory, East Lansing, Michigan, United States) on the campus of Michigan State University (MSU). JENSA enables direct measurements of various hydrogen- and helium-induced astrophysical reactions in inverse kinematics with radioactive, low-energy beams provided by NSCL and, later, with beams available at the Facility of Rare Isotope Beams (FRIB)~\cite{Thoennessen2010NPA,Gade2014NPN,Gade2016PS}. The ReA3 facility can provide ion beam energies between 0.3 and 6.0\,MeV/u.

In a stand-alone operational mode, the JENSA gas-jet target at ReA3 can be surrounded with silicon detector arrays in order to measure the lighter reaction products from \textit{e.\,g.} ($\alpha$,\,p) reactions that play a critical role in the $\alpha$p-process in X-ray bursts~\cite{Cyburt2016APJ}. Simultaneous measurements of ($\alpha$,\,$\alpha$) reactions give the opportunity to test and to improve $\alpha$-optical models potentials for exotic nuclei, which are presently poorly constrained~\cite{Avrigeanu2017PRC}.
With the JENSA gas-jet target, (d,\,p) neutron-transfer reactions in inverse kinematics can be used to extract information about (n,\,$\gamma$) capture reactions and, in some cases by taking advantage of mirror symmetry, (p,\,$\gamma$) cross sections on unstable nuclei of interest for astrophysics. This method is described \textit{e.\,g.} in~\cite{Jones2010N}. Simultaneously measuring (d,\,d) reactions can further constrain the derived capture cross sections.
Proton capture reaction resonance properties of importance for the rp-process in classical Novae~\cite{Jose2001APJ,Iliadis2002APJS} or X-ray bursts~\cite{Thielemann2001PPNP,Parikh2008APJS,Cyburt2016APJ} can also be determined with JENSA using proton scattering, or ($^3$He,\,d) proton-transfer reactions.

JENSA will serve as the gas target for the recoil separator SECAR~\cite{Berg2016NIMB} currently under construction at MSU. In combination with SECAR and additional $\gamma$-ray detectors, JENSA will enable direct measurements of (p,\,$\gamma$) and ($\alpha$,\,$\gamma$) reactions of astrophysical interest in the rp-process in Novae and X-ray bursts, in the synthesis of long lived radioactive isotopes in explosive Si burning in supernovae~\cite{The1998APJ,Magkotsios2010APJS,Iliadis2011APJS}, in the $\nu$p-process in supernova neutrino driven winds~\cite{Frohlich2014JPG}, and in other astrophysical sites.

The JENSA gas-jet target has been optimized to take advantage of low-intensity radioactive beams. The high gas compression of more than 2.5\,MPa provides an unprecedentedly high number density of $\sim\!\!10^{19}$ atoms/cm$^2$. Unlike plastic thin-foil targets or gas targets with windows, the gas-jet target is chemically pure, reducing beam-induced background as well as energy straggling and angular scattering. The excellent reaction localization within a few millimeters enables precise measurements of angular distributions and a good mass separation of recoils from scattered beam when combining JENSA with the SECAR recoil separator. Recirculation of the target gas in JENSA enables the use of expensive gases such as $^{3}$He. The design of the JENSA system minimizes the amount of gas needed to about 160 liters at standard temperature and pressure.

A previous instrumentation paper of JENSA~\cite{Chipps2014NIMA} describes the initial configuration at Oak Ridge National Laboratory (ORNL); here we discuss the implementation at NSCL. The subsequent section describes changes, upgrades, and new features of the JENSA system at NSCL. Then we describe in detail the current configuration of the recirculating gas system at NSCL and present results from nozzle optimization experiments and precise measurements of the target thickness. Finally, we compare JENSA with other supersonic gas-jet targets.

\begin{table*}
	\caption{Measured flow of compressed $^4$He gas and ambient pressure around the gas-jet target for compressor discharge pressures in the range of the measurements. Note that Reference \cite{Chipps2014NIMA} reports over-pressures with respect to atmosphere, as psig in the first row, whereas the present work states absolute pressures, as MPa in the second row.}
	\label{tab:discharge_flow}
	\begin{tabularx}{\textwidth}{lXD{.}{.}{-1}D{.}{.}{-1}D{.}{.}{-1}D{.}{.}{-1}D{.}{.}{-1}D{.}{.}{-1}D{.}{.}{-1}}
		\hline
		\hline
		Compressor discharge pressure
		& (psig)		& 100	& 150	& 200	& 250	& 300	& 350	& 400 \\
		& (MPa)			& 0.79	& 1.14	& 1.48	& 1.83	& 2.17	& 2.51	& 2.86 \\
		\hline
		$^4$He flow in front of the nozzle
		& (dm$^3$/s)	& 4.7	& 6.2	& 7.7	& 9.2	& 10.8	& 12.5	& 14.2 \\
		Ambient pressure around the jet
		& (Pa)			& 5.1	& 6.9	& 8.7	& 10.7	& 13.3	& 15.2	& 17.3 \\
		\hline
		\hline
	\end{tabularx}
\end{table*}

\section{New features of the JENSA system at NSCL}

\begin{figure}
	\centering
	\includegraphics[scale=1]{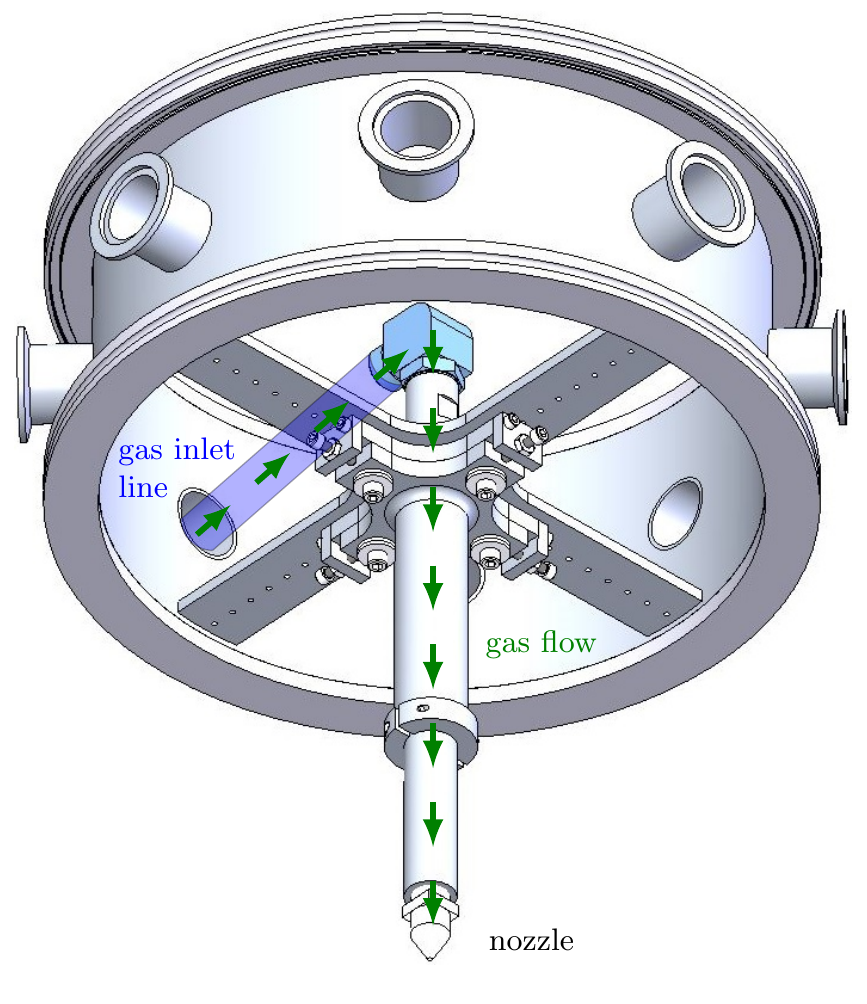}
	\caption{Collar on top of the vacuum chamber that enables precision alignment of the nozzle in the sub-millimeter range and allows the high pressure line to enter the chamber through one of the eight new ports for maximum flexibility. The new mount reduces the solid angle around the target lost to hardware significantly.
	}
	\label{fig:collar}
\end{figure}

The JENSA recirculating gas system at NSCL features an improved pumping scheme (\autoref{sec:TheRecirculatingGasSystem}), where \textit{e.\,g.} the EBARA A10S multi-stage dry vacuum pumps at the final pumping stage before the compressor inlet were replaced with three Leybold DRYVAC DV~650 dry compressing screw vacuum pumps to provide more reliable operation and remove the need for some gas to be bled into the EBARA labyrinth seal. In combination with the pumps, a new water cooler was installed, reducing the gas temperature before the last compression stage. Additionally, a new collar (\autoref{fig:collar})
was installed on top of the vacuum chamber that supports a new nozzle mount equipped with multiple set screws for a precise alignment of the nozzle in all three dimensions to within 0.1\,mm.

In contrast to the original setup at ORNL, the compressor is now located close to the target chamber surrounded by other experimental stations and work places posing particular challenges for the reduction of noise and vibrations. The compressor is installed on top of a concrete base that is elevated by one meter to give additional space for pumps underneath and hence a more compact setup with a minimum total inner volume. To dampen vibrations, the platform is mounted with springs to a support structure of steel. The compressor and its base are enclosed with pads of noise isolating material, reducing the sound pressure level directly next to the compressor from approximately 83.1 to 72.8\,dB(A), corresponding to a reduction of the audible noise level by about a factor of two.

\begin{figure}
	\centering
	\includegraphics[scale=1]{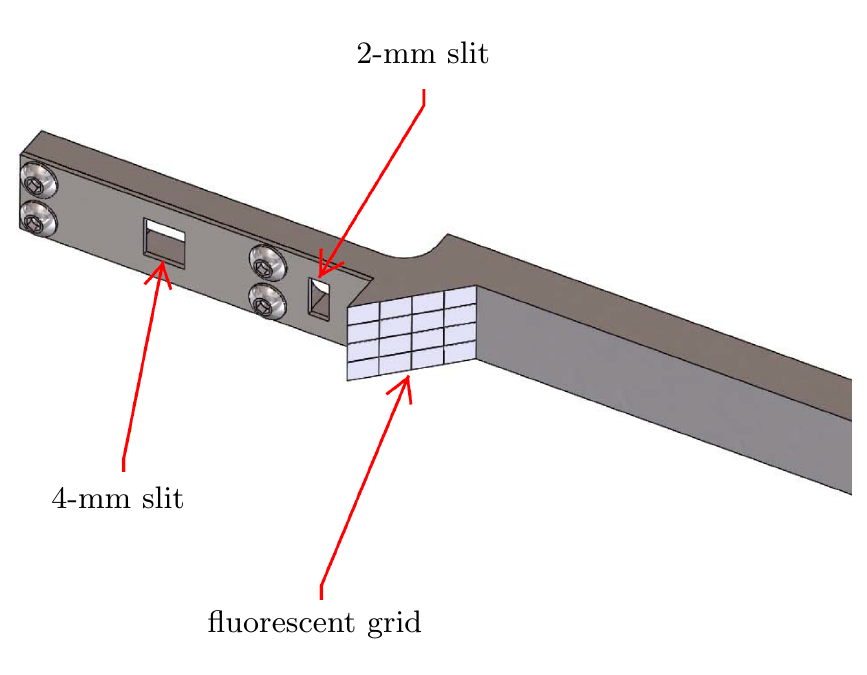} 
	\caption{Computer-aided design of the new JENSA beam viewer, that features a 4$\times$4-mm$^2$ and a 2$\times$4-mm$^2$ slit for transmission studies with low-intensity radioactive ion beams with less than 1000 particles per second, as well as a fluorescent plate featuring a carved grid with 2-mm spacing projected perpendicular to the beam axis. The plate is coated with red CRT phosphor Y2O2S:Eu, that is sensitive to beams with more than 1000 particles per second.}
	\label{fig:viewer}
\end{figure}

Another upgrade is the implementation of a new beam viewer (\autoref{fig:viewer})
at the target location. This addresses the long standing problem of ensuring complete beam overlap when using gas jet targets~\cite[p. 725]{Filippone1986ARNPS}. The viewer can be inserted exactly at the position of the jet, enabling beam tuning to match the jet properties. At instantaneous beam rates of more than 1000 particles per second the beam can be visualized on a fluorescent plate coated with red CRT phosphor Y2O2S:Eu from Nichia (Tokushima, Japan). This plate is $\sqrt{128}$\,mm wide, 8\,mm high and mounted at an angle of 45$^{\circ}$ with respect to the beam axis. The horizontal lines of its carved grid have spaces of 2\,mm, the vertical lines are $\sqrt{8}$\,mm apart, so that the projection of the plate perpendicular to the beam axis is a 4$\times$4 grid of 2\,mm $\times$ 2\,mm cells. For intensities lower than 1000 particles per second, the maximum transition through the center of the gas jet can be verified by transmission studies with and without two 2- and 4-mm wide slits in conjunction with a downstream ionization chamber or other detectors. The viewer is attached to a stepper motor drive outside the chamber with a 50-mm stroke. Tests during alignment demonstrated that the viewer position is reproducible to less than a tenth of a millimeter.

\section{The recirculating gas system at NSCL\label{sec:TheRecirculatingGasSystem}}

\begin{figure*}
	\centering
	\includegraphics[scale=1]{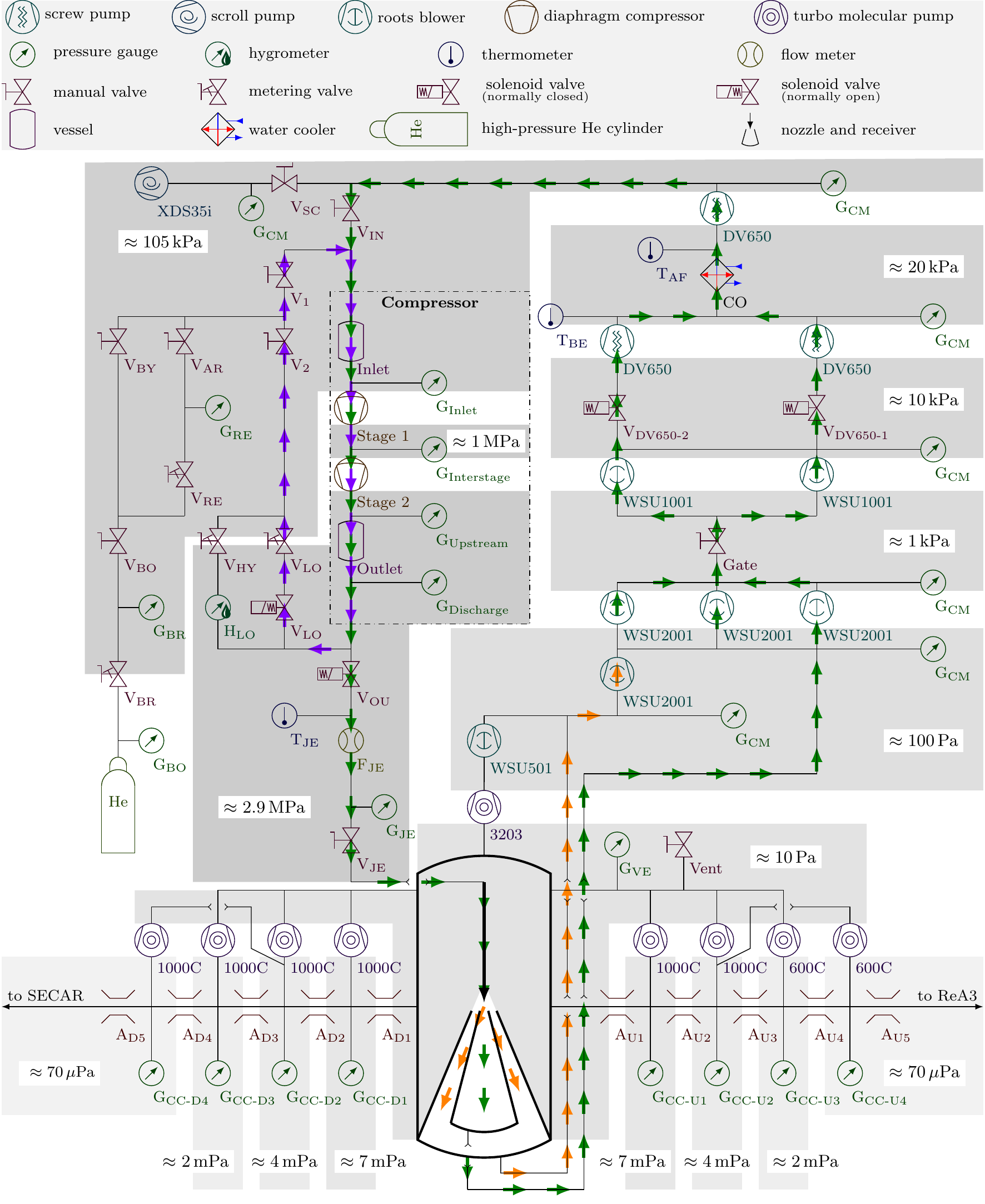}
	\definecolor{c12}{rgb}{.250,.000,.125}
	\caption{Schematic of the JENSA system as installed at NSCL. The major gas flow during operation is indicated with green arrows. The gas flow through the nozzle and hence the jet density is adjustable by needle valve \textcolor{c12}{V$_\text{LO}$} regulating the flow through the loop (violet arrows). Less than 10\% of the gas is caught by the outer receiver (orange arrows). Regions of similar pressure are grouped by light gray shadows. The different kinds of pumps are described in detail in the text.}
	\label{fig:JENSAsystem}
\end{figure*}

A maximum helium gas pressure of about 3\,MPa is provided by a PDC-4-100-500(150) two-stage industrial grade diaphragm compressor (\autoref{fig:JENSAsystem}).
A high pressure storage vessel at the compressor outlet realizes a constant outlet pressure by dampening pressure fluctuations due to the moving diaphragms (see \cite{Chipps2014NIMA} for further details). For example, when the compressor gauge G$_\text{Upstream}$ (\autoref{fig:JENSAsystem}) measured a fluctuating pressure between about 2.8 and 3.0\,MPa, gauge G$_\text{Discharge}$ showed a constant pressure of 2.9\,MPa. To reduce the gas flow through the target, part of the gas can run through a loop back to the compressor inlet (violet arrows in \autoref{fig:JENSAsystem}). This is regulated by the needle valve V$_\text{LO}$ and enables the discharge pressure to be adjusted between 0.79 and 2.86\,MPa. In the loop, the gas can further flow through the probe of a hygrometer H$_\text{LO}$ to measure the humidity of the gas. This device monitors the gas composition to determine if there are any air leaks into the gas system. In addition, the loop is connected to a high pressure gas cylinder that supplies the JENSA system with gas during the startup.

Temperature, flow (\autoref{tab:discharge_flow}), and pressure of the compressed gas are measured (T$_\text{JE}$, F$_\text{JE}$ and G$_\text{JE}$, respectively) directly before the compressed gas enters the target chamber. A feed through guides the high pressure pipe into the vacuum chamber, where a flexible Swagelok pipe connects to the nozzle holder mounted at the collar shown in \autoref{fig:collar}. Several nozzles with different designs are available to be mounted and described in detail in \autoref{sec:AvailableNozzles}. Eventually, the gas is further compressed by the shape of the nozzle and passes through the vacuum chamber as a supersonic jet.

Residual gas drifting through the vacuum chamber into the beam line is evacuated by two sets of differential pumping stages with water-cooled turbo molecular pumps. The set upstream of the vacuum chamber consists of two TURBOVAC 1000~C and two TURBOVAC 600~C pumps with a maximum pumping speed for helium of 1.0 and 0.6\,m$^3$/s, respectively. The downstream set consists of four TURBOVAC 1000~C pumps to provide sufficient pumping for the vacuum requirements for the SECAR recoil separator~\cite{Berg2016NIMB}. The pumps reduce the pressure along the beam lines down to the $10^{-5}$\,Pa range and pump the gas back into the main vacuum chamber. That gas, together with residual gas from the jet drifting into the vacuum chamber, is pumped by a Shimadzu TMP-3203LMC magnetically levitated turbo molecular pump that has a maximum pumping speed of 3\,m$^3$/s for helium and that is backed by a RUVAC WSU501 roots pump with a speed of 140\,dm$^3$/s. This keeps the pressure in the main vacuum chamber well below 10\,Pa.

The gas jet is caught by a two-tier receiver system with an outer and inner cone-shaped catcher with increasing diameters. The outer part of the jet (less than 10\% of the total gas flow) goes into the outer receiver (light yellow arrows in \autoref{fig:JENSAsystem}) and is pumped by a RUVAC WSU2001 roots pump with a maximum pumping speed of 570\,dm$^3$/s. The major part of the jet is pumped through the inner receiver.

Residual gas from the vacuum chamber (pumped by the Shimadzu and the RUVAC WSU501), the gas from the outer jet (pumped by the RUVAC WSU2001), and gas from the inner jet are combined and subsequently pumped by three additional RUVAC WSU2001 roots pumps. At the inlet of those pumps, a pressure of about 100\,Pa was measured. Those three RUVAC WSU2001s are in turn backed by two RUVAC WSU1001 roots pumps with a maximum pumping speed of 280\,dm$^3$/s each, compressing the gas further to about 1\,kPa. The two final pumping stages before the compressor inlet are realized by three DRYVAC DV~650 dry compressing screw vacuum pumps with a maximum pumping speed of 181\,dm$^3$/s each. The first two DV650 pumps compress the gas to about 20\,kPa, before its temperature is reduced by a water cooler from about 310\,K at T$_\text{BE}$ to about 285\,K at T$_\text{AF}$ (\autoref{fig:JENSAsystem}). The third DV650 pump increases the pressure of the gas to about 105\,kPa since the minimum inlet pressure of the compressor has to be slightly higher than 1\,atm.

\section{Available nozzles \label{sec:AvailableNozzles}}

Creating a suitable supersonic gas jet highly depends on the geometry of the nozzle. For the JENSA gas-jet target, six different nozzles labeled A to F are available (\autoref{tab:nozzles}). Their design is based on de Laval \cite{DeLaval1894USP} shaped nozzles and associated detailed studies \cite{Bittner1979NIM,Schmid2012RSI}.
\begin{table}
	\caption{Characteristics of the nozzles used for jet thickness measurements. The six nozzles differ in diameter of the neck (the minimum inner diameter) $d_\text{neck}$, the inner diameter at the end of the exhaust $d_\text{exhaust}$, the length of the exhaust from the neck to the the end $l_\text{exhaust}$, the length of the nozzle pipe $l_\text{pipe}$ and the length of the nozzle tip $l_\text{tip}$. A schematic drawing of the nozzles is shown in \autoref{fig:nozzles}. Tolerances are stated in the last row and dimensions are given in mm.}
	\label{tab:nozzles}
	\begin{tabularx}{\columnwidth}{X*{5}{D{.}{.}{-1}}}
		\hline
		\hline
		Nozzle		& \multicolumn{1}{c}{$d_\text{neck}$}
						 	& \multicolumn{1}{c}{$d_\text{exhaust}$}
						 			& \multicolumn{1}{c}{$l_\text{exhaust}$}
						 					& \multicolumn{1}{c}{$l_\text{pipe}$}
								 					& \multicolumn{1}{c}{$l_\text{tip}$} \\
		\hline
		A			& 0.80	& 2.39	&  7.62	& 32.64	& 14.43 \\
		B			& 0.80	& 3.20	& 11.40	& 31.39	& 13.72 \\
		C			& 0.90	& 2.69	&  8.61	& 31.65	& 14.17 \\
		D			& 1.00	& 3.00	&  9.50	& 30.76	& 13.89 \\
		E, F		& 1.10	& 3.30	& 10.49	& 29.77	& 13.64 \\
		\hline
		Tol.		& 0.08	& 0.08	&  0.25	&  0.25	&  0.25 \\
		\hline
		\hline
	\end{tabularx}
\end{table}
\begin{figure}[t!]
	\centering
	\includegraphics[scale=1]{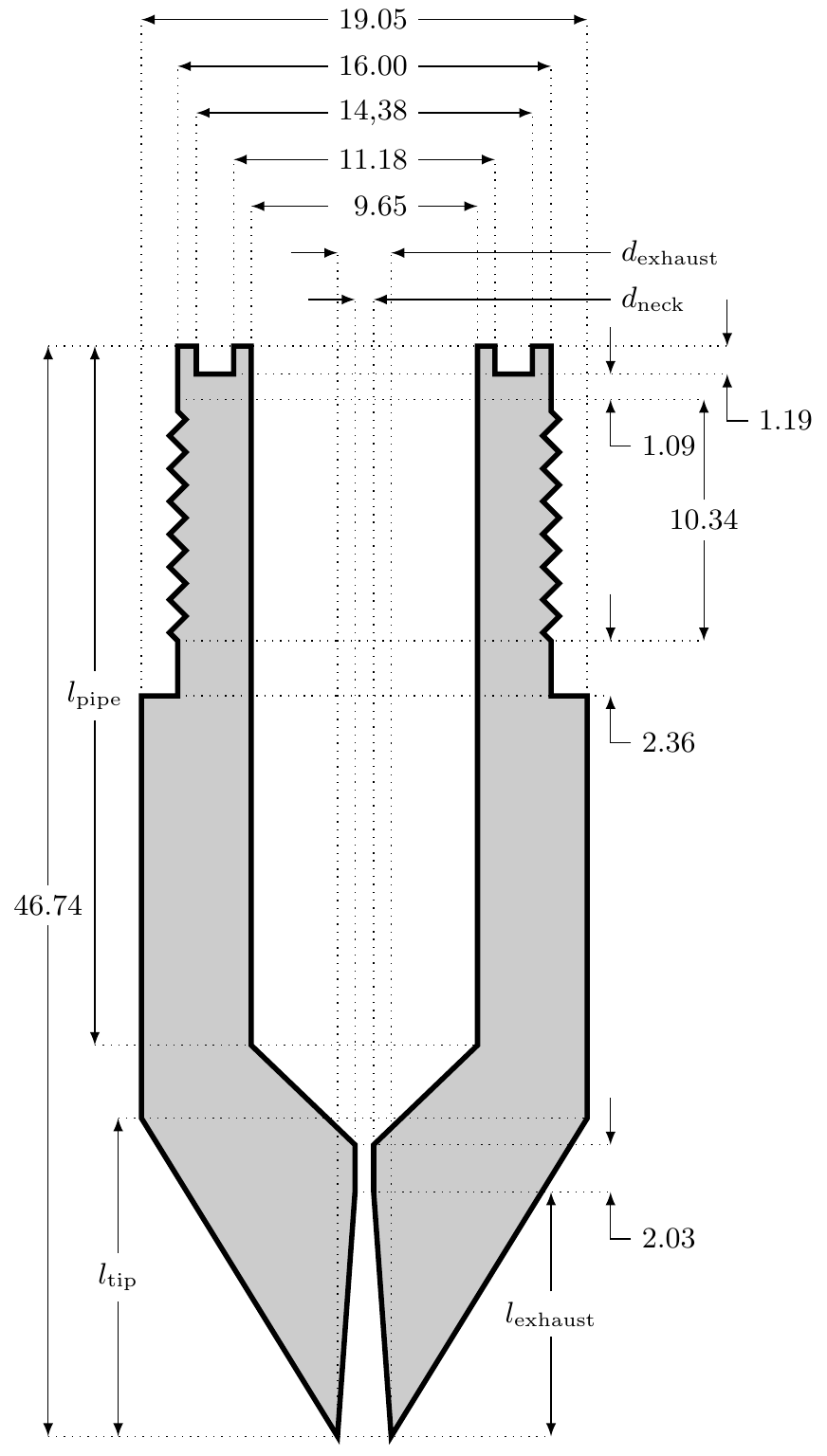} 
	\caption{Schematic drawing of the nozzles used for the jet thickness studies. Dimensions are in mm and the nozzle is illustrated with a magnification of about 2.5. For the variable lengths ($l_\text{pipe}$, $l_\text{tip}$, and $l_\text{exhaust}$) and diameters ($d_\text{neck}$ and $d_\text{exhaust}$), see \autoref{tab:nozzles}.}
	\label{fig:nozzles}
\end{figure}

All nozzles in the present work have a cylindrical symmetry (\autoref{fig:nozzles}); a total length (from the o-ring groove to the end of the exhaust) of 46.74\,mm; a maximum outer diameter of 19.05\,mm; a maximum inner diameter of 9.65\,mm; the groove for o-rings with an inner diameter of 11.18\,mm, a width of 3.2\,mm and a depth of 1.19\,mm; and a 10.34-mm long 3/4-16 thread with a diameter of 16\,mm starting after 2.38 mm from the upper end (with the o-ring groove). In order to tighten the nozzle into its holder with tools, a 6.35-mm high block with a square cross section of 19.05\,mm was added 2.36\,mm below the thread. This is also the outer diameter of the cylindrical shaped body of the nozzle below that block. Outside, all nozzles have a cone-shaped tip at the end with a lateral-surface angle of $30^\circ$ with respect to the jet axis. Inside, the cylindrically shaped pipe changes into a cone with a lateral-surface angle of $45^\circ$, then into a 2.03-mm long neck and finally into another cone shaped exhaust with a total opening angle of $12^\circ$. The nozzles differ in the lengths of the cylindrical-shaped pipe, the tip and the exhaust, as well as in the diameter of the neck and of the exhaust opening at the end of the nozzle. All those distances are shown in \autoref{fig:nozzles} and listed in \autoref{tab:nozzles}.

Nozzles used in this work are similar to the de\,Laval nozzles developed for steam turbines~\cite{DeLaval1894USP}. However, G.\,P. de\,Laval designed smooth shapes whereas we used edges (\autoref{fig:nozzles}) that can be characterized and precisely fabricated more easily.

\section{Jet thickness studies\label{sec:JetThicknessStudies}}

The thickness of the gas jet for the six available nozzles (see \autoref{sec:AvailableNozzles}) was measured by the energy loss of 5.5-MeV $\alpha$ particles. An activity-standard $^{241}$Am source was mounted inside the vacuum chamber facing a Micron style BB15 silicon strip detector~\cite{Bardayan2013NIMA}, with the jet positioned between both, 26\,mm from the source and 20\,mm from the detector. In front of the $^{241}$Am source, a $0.64$-mm thick aluminum sheet with a centered bore-hole of about $0.5$\,mm diameter was mounted as collimator so that the active area was small compared to the jet thickness.

\begin{figure}
	\centering
	\includegraphics[scale=1]{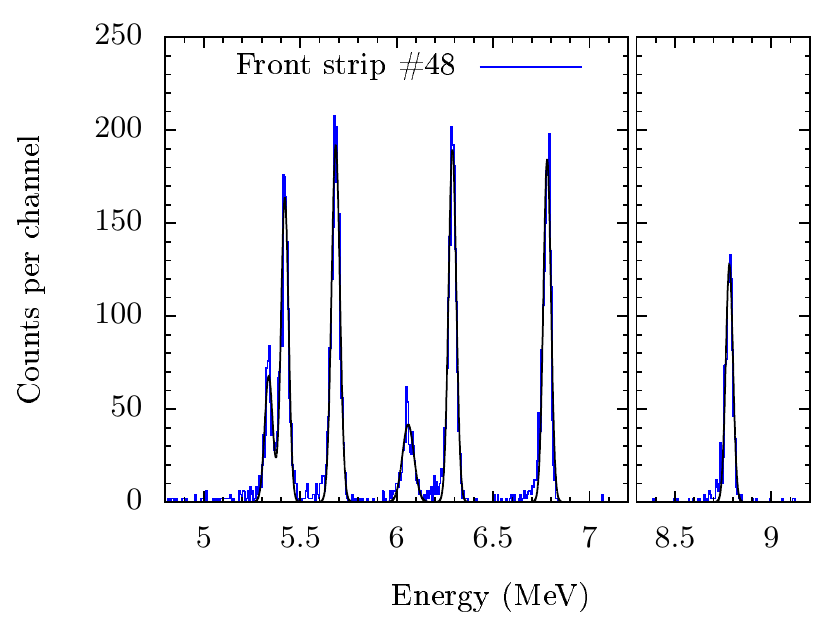}
	\caption{Spectrum from an activity standard of $^{228}$Th measured by front strip \#48 of the BB15 detector. Gaussian functions (black curves) are fitted to the detected $\alpha$ counts per channel (blue histogram), including peaks at (from left to right) 5340.36, 5423.15 (both from the decay of $^{228}$Th), 5685.37 ($^{224}$Ra), 6050.78 ($^{212}$Bi), 6288.33 ($^{220}$Rn), 6778.3 ($^{216}$Po) and 8784.86\,keV ($^{212}$Po). Empty channels between 7.2 and 8.3\,MeV are skipped. With similar spectra for each of the 64~front strips, the region of the energy loss measured with an activity standard of $^{241}$Am (\autoref{fig:americium}) has been calibrated by interpolating the known peaks.}
	\label{fig:thorium}
\end{figure}

\begin{figure}
	\centering
	\includegraphics[scale=1]{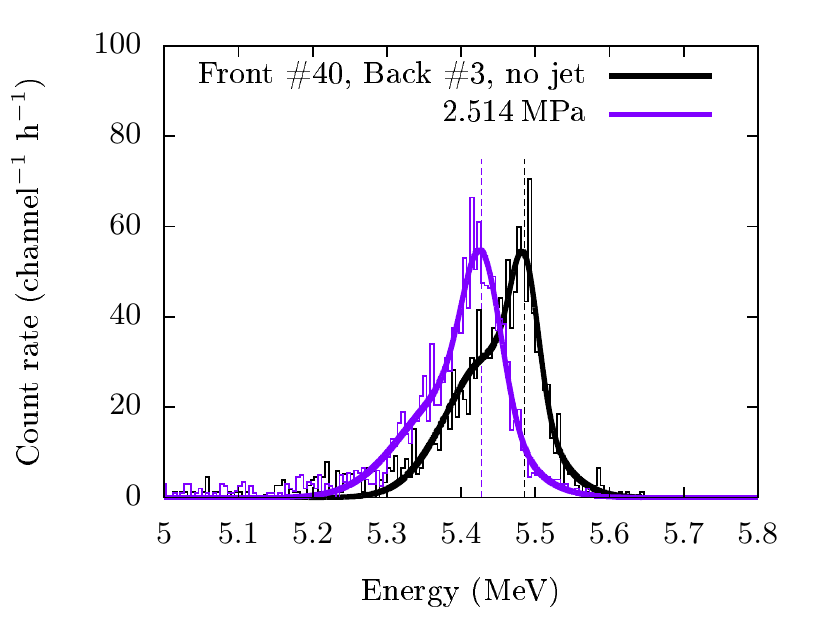}
	\caption{Spectrum from an activity standard of $^{241}$Am measured by front strip \#40 in coincidence with back strip \#3. The $\alpha$ particles were measured without (black histogram) and with a jet at an input pressure of 2.514\,MPa (purple histogram). Thick curves are double Gaussian functions fitted to the histograms using the nonlinear least-squares Marquant-Levenberg algorithm~\cite{Levenberg1944QAM,Marquardt1963JSIAM}. The high-energy peaks correspond to $\alpha$ particles leaving the collimator in front of the source with the full energy of 5485.56-keV. The energy difference of the fitted maximums (dashed vertical lines) correspond to an energy loss of $57.5\pm0.8$\,keV in the jet. The low-energy shoulders of both peaks result from energy losses at the edge of the collimator bore-hole and from another $^{241}$Am decay emitting $\alpha$ particles with 5442.8 keV. Note that similar spectra have been analyzed for a total of 256 pixels $\times$ 6~nozzles $\times$ 3 input pressures. The determined energy losses are shown in \autoref{fig:eloss}.}
	\label{fig:americium}
\end{figure}

The BB15 is a highly segmented double sided silicon strip detector with 64~vertical strips (each 1172.5\,$\mu$m wide and 40.3\,mm high) at the front side and four horizontal strips (each 75.0\,mm wide and 10087.5\,$\mu$m high) at the back side. Each of the 64 front strips of the BB15 detector is oriented parallel to the jet axis and energy calibrated with an activity-standard $^{228}$Th source facing the BB15 detector. The resulting $\alpha$ spectrum is shown in \autoref{fig:thorium}.
With this energy calibration, the energy of the $^{241}$Am $\alpha$ particles with and without the jet (\autoref{fig:americium})
was measured at compressor discharge pressures of 1.136, 1.825 and 2.514\,MPa for the six different nozzles listed in \autoref{tab:nozzles}.

\begin{figure*}
	\centering
	\includegraphics[scale=0.98]{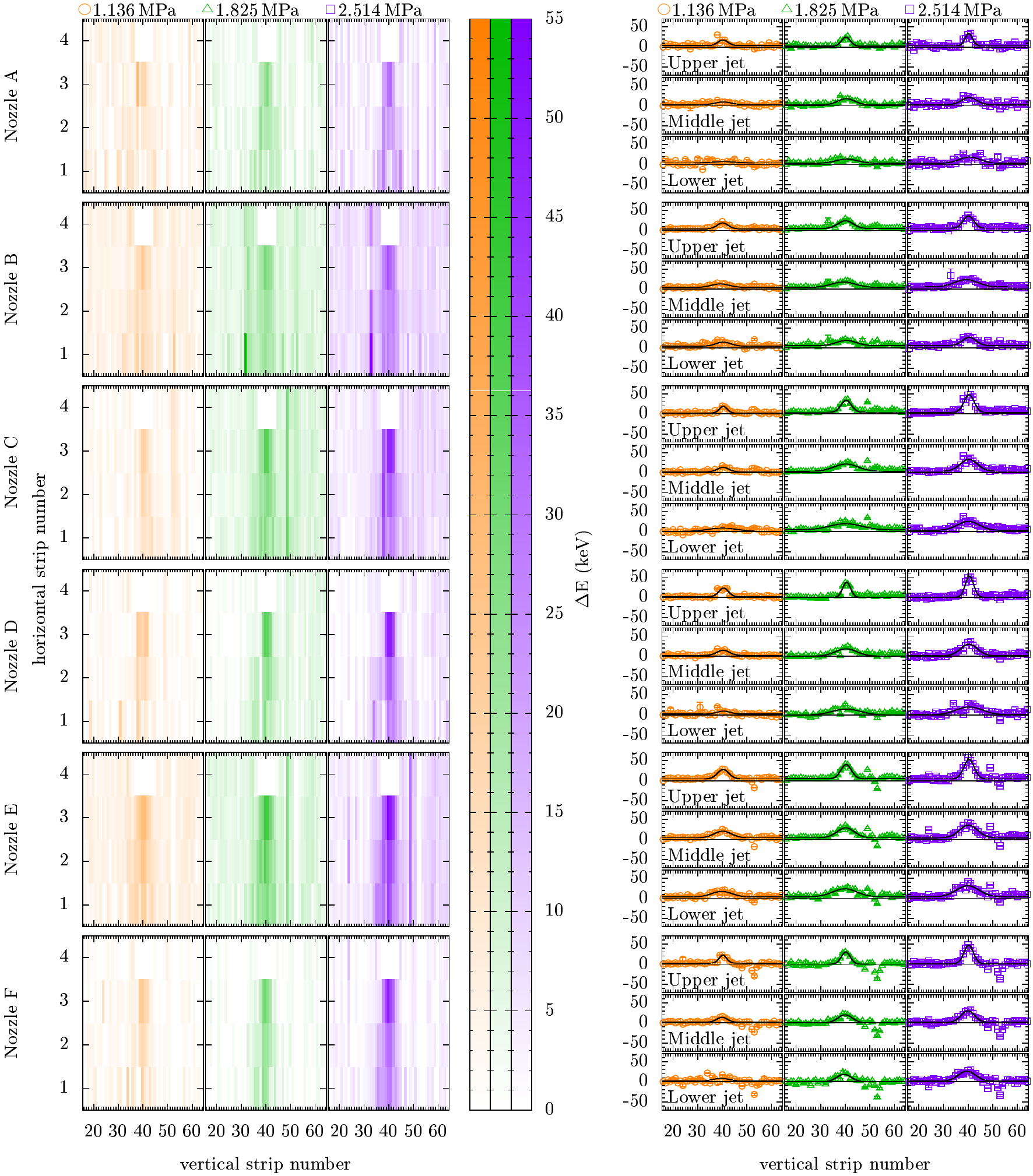}
	\caption{Energy loss of 5486-keV $\alpha$ particles crossing through the upper, middle and lower (horizontal strip number 3, 2 and 1) part of the jet at compressor discharge pressures of 1.136\,MPa (orange), 1.825\,MPa (green) and 2.514\,MPa (purple) of six different nozzles (\autoref{tab:nozzles}). The upper part indicates the first 4\,mm of the jet, the middle part 4 to 8\,mm and the lower part 8 to 12\,mm. The energy loss for all jet segments at a single pressure is shown in each panel of the left three columns. The right three columns show the energy loss for a single jet segment at three different pressures. Horizontal strip number 4 is partly shadowed by the nozzle (empty squares at vertical strip numbers 37 to 44 in the left panels) and hence not shown in the right panels. Solid black lines lines are Gaussian functions fitted to the measured energy losses per vertical strip number using the nonlinear least-squares Marquardt-Levenberg algorithm~\cite{Levenberg1944QAM,Marquardt1963JSIAM}. Note that the first three rows of the right three columns have been measured with Nozzle A, rows 4 to 6 with nozzle~B and so on. The ambient pressure in the surrounding vacuum chamber is given in \autoref{tab:discharge_flow}. The negative energy-loss artifacts for the measurements of nozzle~E and F are caused by the noisy front strip \#53, that is not influencing the fit region of the peak.}
	\label{fig:eloss}
\end{figure*}

\begin{figure*}
	\centering
	\includegraphics[scale=1]{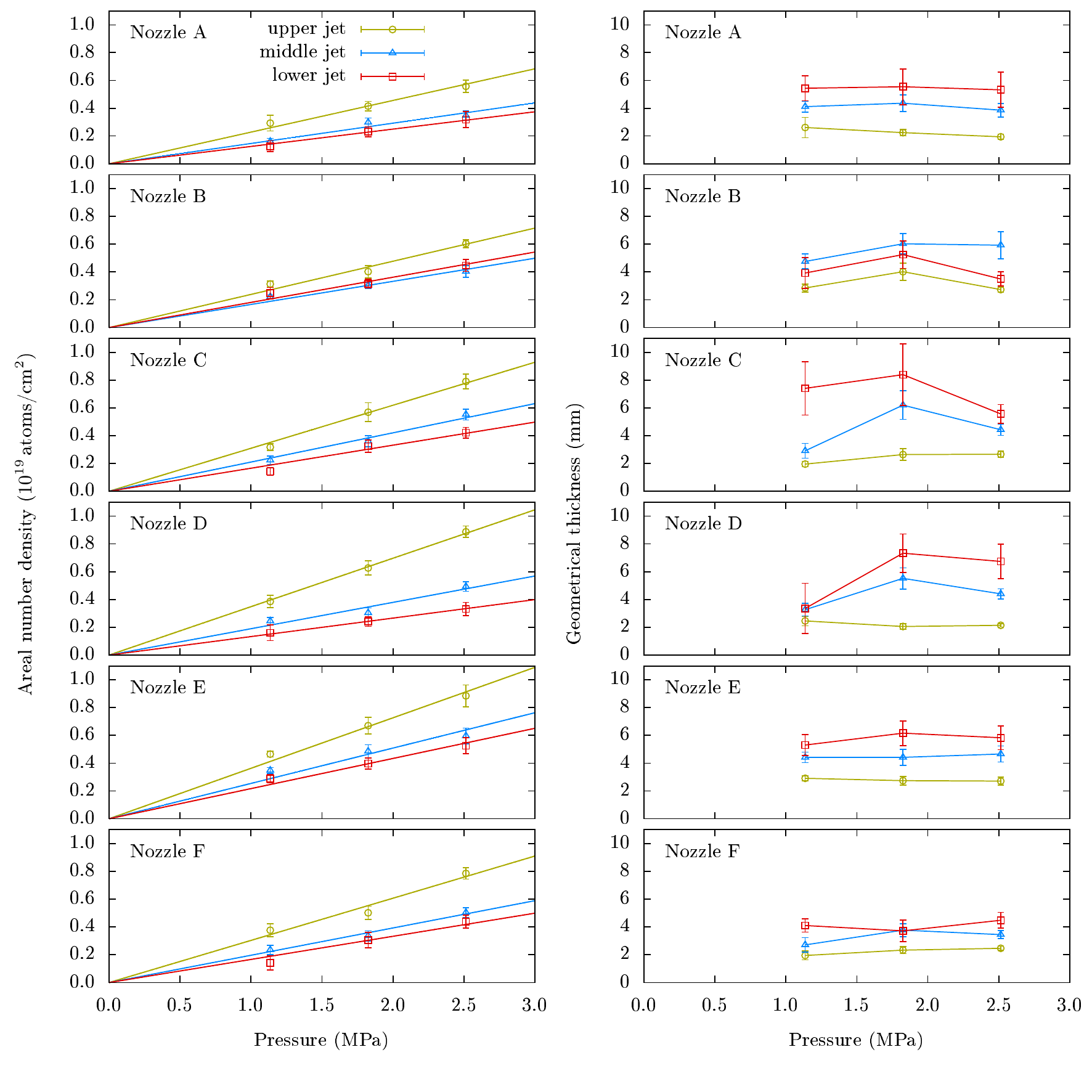}
	\caption{Results of the jet thickness measurements for each of the six nozzles. The number of atoms per area (left panels) has been calculated from the energy loss using a measured~\cite{Hanke1978NIM} stopping power of $(5.973\pm0.018)\times10^{-15}$\,eV\,cm$^2$/atoms at an $\alpha$ energy of $(5485.56\pm0.12)$\,keV. Solid lines are linear functions fitted to the areal densities determined in \autoref{fig:eloss} and forced to intercept the origin. The geometrical jet widths (right panels) have been determined by comparing the FWHMs from the energy loss distributions per detector strip (see \autoref{fig:eloss}) with the number of shadowed detector strips caused by cylinders with different well-known thicknesses placed at the jet position. For the geometrical jet widths, solid lines are only given to guide the eye in order to show that the upper part of the jet is nearly independent from the nozzle-input pressure. This is indicated by nearly horizontal dark-yellow lines. The small uncertainties for the upper part of the jet indicate further that the jet is best defined directly after it leaves the nozzle. Note that observed differences between the identically designed nozzles~E and F (\autoref{tab:nozzles}) are likely to be caused by minor differences in the course of fabrication. Hence, an additional uncertainty of up to 17\% has to be considered on the results that can be obtained with a nozzle with identical nominal dimensions.}
	\label{fig:thicknesses}
\end{figure*}

Measuring the energy deposited by $\alpha$ particles in each front strip in coincidence with each back strip, gives a total of 256~pixels. Hit patterns in the left panels of \autoref{fig:eloss}
show that the inner pixels of the uppermost horizontal strip are partially shadowed by the nozzle. No signals from $\alpha$ particles in the vertical front strips \#37 to \#44 are detected in coincidence with the horizontal back strip \#4. Hence, the energy loss as a function of the front strip number is only given for back strips 1 to 3 and referred to as lower (8 to 12\,mm from the nozzle), middle (4 to 8\,mm from the nozzle) and upper (within 4\,mm below the nozzle) part of the jet in the right panels of \autoref{fig:eloss} and when displaying the fit parameters in \autoref{fig:thicknesses}.

When the supersonic gas jet leaves the nozzle, the gas is expanding into the vacuum chamber. This causes the density to decrease with increasing distance from the nozzle. This is the case for all six nozzles, as shown in \autoref{fig:eloss}. The figure also shows that the jet from all six nozzles is causing a higher energy loss and hence getting denser with increasing compressor discharge pressure, caused by the higher gas throughput. This increase can be described by a linear function as shown in the left panels of \autoref{fig:thicknesses}.

In order to determine the geometrical width of each jet, the measurement was calibrated using four small cylinders of different well-known diameters that were placed at the location of the jet. The diameter of a cylinder and the number of shadowed strips are directly proportional. Applying this geometrical calibration to the FWHMs shown in \autoref{fig:eloss} resulted in the geometrical jet widths shown in the right panels of \autoref{fig:thicknesses}. We find that the jet width is less dependent on the pressure in the upper part of the jet near the nozzle.

The maximum areal density of $(9\pm3)\times10^{18}$\,atoms/cm$^2$ was measured in the upper part of the jet with nozzle~D at an input pressure of 2.86\,MPa. A slightly lower jet thickness was measured in the upper jet with nozzle~E. For this nozzle the areal densities in the middle and lower part of the jet were determined to be more than $10^{18}$\,atoms/cm$^2$ higher than for nozzle~D. The geometrical width for jets from nozzle~E also show much less dependence on the input pressure (\autoref{fig:thicknesses}) and hence, the overall best jet characteristics were achieved with nozzle~E. Comparing the parameters in \autoref{tab:nozzles}, we find that the broad neck together with the long exhaust in nozzle~E are the best combination to achieve a homogeneous jet with high density.

The geometrical width determined in this work (\autoref{tab:literature}) is significantly smaller than previously measured at ORNL. Chipps \textit{et al.}~\cite{Chipps2014NIMA} studied the geometrical thickness of the jet with the same nozzles~A and B as listed in \autoref{tab:nozzles}. But the width of the present work stated in \autoref{tab:literature} refers to nozzle~D that generated the most dense jet. Nozzle D has a longer exhaust in combination with a broader neck while maintaining the maximum width of the exhaust. In addition, the BB15 detector was aligned closer to the jet resulting in a better spatial resolution.

\begin{table*}
	\caption{Review of supersonic helium jet targets for nuclear physics similar to the JENSA approach. The quoted values were obtained in the given distance from the nozzle. Results from this work were measured with nozzle~D.}
	\label{tab:literature}
	\begin{tabularx}{\textwidth}{lXD{.}{.}{-1}D{.}{.}{-1}@{$\;\pm\;$}lD{.}{.}{-1}@{$\;\pm\;$}lr@{\,}lc}
		\hline
		\hline
					& 		& \multicolumn{1}{c}{Input pressure}
									& \multicolumn{2}{c}{$^4$He jet density}
													& \multicolumn{2}{c}{$^4$He jet FWHM}
																	& \multicolumn{2}{c}{Distance from nozzle}
																					& \\
		Location	& Year	& \multicolumn{1}{c}{(kPa)}
									& \multicolumn{2}{c}{($10^{18}$\,at./cm$^2$)}
													& \multicolumn{2}{c}{(mm)}
																	& \multicolumn{2}{c}{(mm)}
																					& Reference \\
		\hline
		M{\"u}nster	& 1982	& 200	& 0.34	& 0.06	& 2.5	& 0.2	& \multicolumn{2}{c}{1 to 5}
																					&~\cite{Becker1982NIM}	\\
		Stuttgart	& 1991	& 38	& 0.078	& 0.013	& 2.6	& 0.2	& $\sim$ & 1.5	&~\cite{Griegel1991}	\\
		Notre Dame	& 2012	& 150	& 0.259	& 0.021	& 2.2	& 0.2	& $\sim$ & 4	&~\cite{Kontos2012NIMA}	\\
		Oak Ridge	& 2014	& 2859	& 10.2	& 0.9	& 5.1	& 0.3	& $\sim$ & 1	&~\cite{Chipps2014NIMA}	\\
		Caserta		& 2017	& 700	& 1.97	& 0.21	& \multicolumn{2}{c}{not reported}
																	& $\sim$ & 5.5	&~\cite{Rapagnani2017NIMB}	\\
		\hline
		East Lansing & 2018	& 2515	& 9.0	& 0.3	& 2.03	& 0.09	& \qquad\quad $\lesssim$
																	 		 & 4	& this work	\\
		\hline
		\hline
	\end{tabularx}
\end{table*}

\section{Comparison with other supersonic gas-jet targets in nuclear physics}

The first gas-jet target for nuclear physics was operated with compressed hydrogen~\cite{Becker1954ZNA}. Becker and Bier report a maximum particle flux of $3.6 \times 10^{17}$\,atoms/(cm$^2$\,s) at a full width at one-fifth maximum (FWFM) of 9.2\,mm, corresponding to a full width at half maximum (FWHM) of 6.0\,mm with a de\,Laval-shaped nozzle~\cite{DeLaval1894USP} and an input pressure of 10.7\,kPa. With a stated average particle velocity of $2.8 \times 10^{5}$\,cm/s and the FWHM as geometrical target thickness, their reported flux corresponds to an areal particle density of $7.8 \times 10^{11}$\,atoms/cm$^2$.

A gas-jet target operated with nitrogen is reported by Tietsch \textit{et al.}~\cite{Tietsch1979NIM}. Based on electron-beam attenuation, electron-beam fluorescence, and large-angle single-scattering of an electron beam, a homogeneous nitrogen gas jet with a maximum thicknesses of $(170\pm9)\,\mu$g/cm$^2$, corresponding to $(7.3\pm0.4) \times 10^{18}$\,atoms/cm$^2$, has been determined. They measured a jet diameter of about 3\,mm produced by an axially symmetric de\,Laval nozzle and an input pressure of 1.3\,MPa.

Bittner \textit{et al.}~\cite{Bittner1979NIM} investigated supersonic jets operated with argon, nitrogen and hydrogen. They report maximum densities for inhomogeneous jets with widths increasing from about 1 to 10\,mm due to their cylindrical and conical shaped nozzles. Thus, with the disadvantage of low density uniformity of the jet, they reached record densities of $(1.60\pm0.16)$\,mg/cm$^2$ corresponding to $(24.1\pm2.4) \times 10^{18}$\,atoms/cm$^2$ for Ar at an input pressure of 2.0\,MPa, $(1.10\pm0.11)$\,mg/cm$^2$ corresponding to $(47\pm5) \times 10^{18}$\,atoms /cm$^2$ for N at 5.0\,MPa and $(30\pm3)\,\mu$g/cm$^2$ corresponding to $(17.9\pm1.8) \times 10^{18}$\,atoms/cm$^2$ for H at 2.0\,MPa.

The first helium jet for nuclear physics was developed by Becker \textit{et al.}~\cite{Becker1982NIM} for $\gamma$-ray spectroscopy measurements. Their recirculating gas target system provided a 2.5-mm wide $^4$He jet at an input pressure of 200\,kPa with a thickness of $0.34 \times 10^{18}$\,atoms/cm$^2$ as determined by proton elastic scattering; excitation functions of narrow resonances in $^{14}$N(p,\,$\gamma$), $^{18}$O(p,\,$\gamma$), $^{20,21,22}$Ne(p,\,$\gamma$), $^{15}$N($\alpha$,\,$\gamma$), and $^{20}$Ne($\alpha$,\,$\gamma$); and the energy shift of a narrow resonance in $^{13}$C($\alpha$,\,n). The target thickness stated in \autoref{tab:literature} results from the average of those three methods.

Another helium-jet target produced by RHINOCEROS through a de\,Laval nozzle is reported by Griegel \textit{et al.}~\cite{Griegel1991}. They do not report an experimentally measured target thickness, but based on their calculated volume density of $3.0 \times 10^{17}$\,atoms/cm$^3$ and a jet width of $(2.60\pm0.20)$\,mm, their thickness was estimated to be $(7.8\pm1.3) \times 10^{16}$\,atoms
/cm$^2$ at 38\,kPa input pressure.

Developed for the St. George Recoil Mass Separator~\cite{Couder2008NIMA,Meisel2017NIMA}, the HIPPO (HIgh Pressure POint-like) gas target~\cite{Kontos2012NIMA} was characterized by
$^{20}$Ne($\alpha$,\,$\alpha$)
elastic scattering measurements and energy loss measurements for a 1.68-MeV $\alpha$-beam through the helium jet. At an input pressure of 150\,kPa before the de\,Laval nozzle, a $(2.20\pm0.20)$-mm wide helium jet with $(2.59\pm0.21) \times 10^{17}$\,atoms/cm$^2$ (as weighted arithmetic mean from both methods) was achieved. Ensuing computational fluid dynamics (CFD) simulations~\cite{Meisel2016NIMA} were found to be consistent with the scattering experiment.

SUGAR, a supersonic gas-jet target for low energy nuclear reaction studies, features a rectangular nozzle and is described by Favela \textit{et al.}~\cite{Favela2015JPCS,Favela2015PRST}. Air, argon and nitrogen jets were characterized with Elastic Backscattering Spectrometry (EBS) and resulted in maximum jet thicknesses of $(1.80\pm0.18)\times10^{18}$\,atoms/cm$^2$ for argon at an input pressures of 300\,kPa and $(3.7\pm0.4)\times10^{18}$\,atoms/cm$^2$ for nitrogen at an input pressure of 250\,kPa. Values are taken from Figure~3 of reference~\cite{Favela2015JPCS} assuming a typographical mistake in the x-axis label, accidentally stating psi instead of bar~\cite{Favela2015PRST}. The areal density of the air jet is only given to be around $10^{18}$\,atoms/cm$^2$ while injecting air at atmospheric pressure. Geometrical widths of the jets are not reported for any target gas.

The latest supersonic He-jet target for nuclear physics was developed for the European Recoil mass separator for Nuclear Astrophysics (ERNA)~\cite{Rogalla1999EPJA}. Rapagnani \textit{et al.}~\cite{Rapagnani2017NIMB} report a jet thickness of $(1.97\pm0.21) \times 10^{18}$\,atoms/cm$^2$ at an input pressure of 0.7\,MPa before the de\,Laval nozzle exploiting the energy loss of a 1-mm collimated $^{12}$C beam. Jet thicknesses based on CFD simulations agree with experimental data within 38\%. The geometrical width of the jet is not quoted. 

Not discussed here are gas-jet targets deployed in high energy nuclear physics~\cite{Shapira1985NIMA,Shapira1985NIMB}, targets for interaction with high-intensity ultra-short laser pulses~\cite{Hosokai2002PEPAC,Schmid2012RSI}, or storage rings~\cite{Gruber1989NIMA,Zapfe1995RSI,Schmidt1997HI} for nuclear astrophysics~\cite{Glorius2017JPCS} where significantly lower areal densities of about 10$^{11}$\,atoms/cm$^2$ are sufficient.

For JENSA at NSCL, a target thickness of $(9.0\pm0.3)\times10^{18}$\,atoms/cm$^2$ and a geometrical target width of $(2.03\pm0.09)$\,mm at an input pressure of 2.5\,MPa was determined from the energy loss of 5.5-MeV $\alpha$ particles from an $^{241}$Am activity standard (\autoref{sec:JetThicknessStudies}). This highly dense jet could be achieved with a high inlet pressure and an optimized inner shape of the nozzle (\autoref{sec:AvailableNozzles}).

\section{Discussion and outlook}

\begin{figure}
	\centering
	\includegraphics[scale=1]{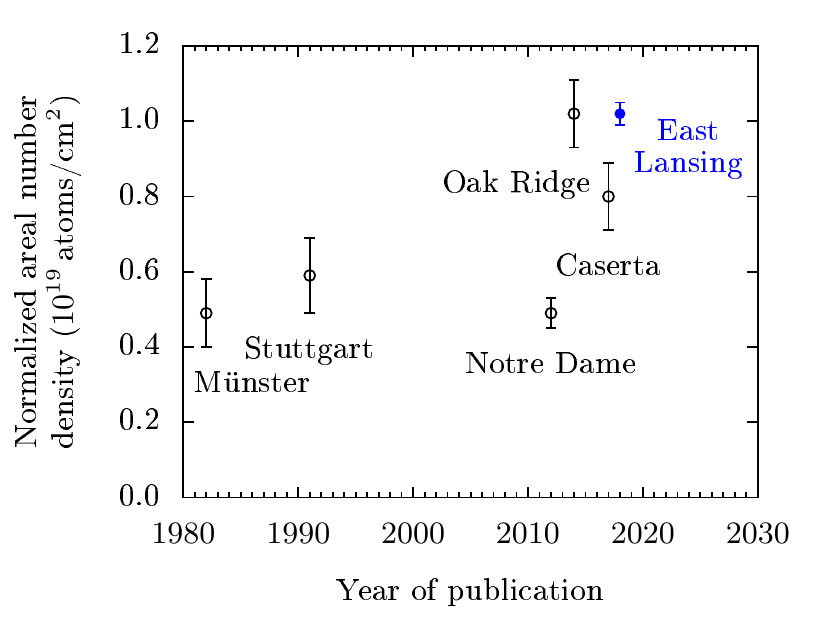}
	\caption{Comparison of helium jet thicknesses from literature (black open circles) and from this work (filled blue circle) normalized to 2.86\,MPa. Areal number densities are scaled by the corresponding input pressure listed in \autoref{tab:literature} because in good approximation, both quantities are proportional. The present result from nozzle~D confirms the jet density achieved during commissioning at ORNL~\cite{Chipps2014NIMA}.}
	\label{fig:literaturedensities}
\end{figure}

In order to compare supersonic helium jets, the areal densities given in \autoref{tab:literature} have been normalized to the maximum jet-input pressure of 2.86\,MPa assuming a linear proportionality between thickness and pressure (\autoref{fig:literaturedensities}).
This assumption is supported by the linear functions shown in the left panels of \autoref{fig:thicknesses}. The current JENSA system at NSCL is able to achieve target thicknesses of about $10^{19}$\,atoms/cm$^2$ and is similar to what was achieved previously at ORNL with nozzles~A and B \cite{Chipps2014NIMA}. The JENSA gas-jet target at NSCL has the highest helium densities of any gas jet target.

Another important figure of merit is the amount of residual gas (\autoref{tab:discharge_flow}) that an ion beam is passing through outside the jet compared to the gas inside the target. For JENSA at an input pressure of 2.5\,MPa, a beam experiences $\sim 7\times10^{16}$\,atoms/cm$^2$ before and $5\times10^{16}$\,atoms/cm$^2$ after the jet, corresponding to a total ratio of outside and inside the jet of less than 1.4\%. This ratio can be decreased even further by reducing the gap between the nozzle and the receiver from 12\,mm, as it was the case for the jet-thickness studies described in \autoref{sec:JetThicknessStudies}, to 6\,mm. This is feasible since, with the help of the new JENSA beam viewer (\autoref{fig:viewer}), the diameter of ion beams provided by ReA3 (2-MeV/u $^1$H; 1.7-MeV/u $^{34}$Ar, $^{39}$K, and $^{40}$Ar beams with full transmission through the 2-mm slit) has been demonstrated to be less than 2\,mm. A reduced gap will decrease the amount of residual gas drifting into the vacuum chamber because the geometry of the jet is much better confined immediately after it leaves the nozzle (\autoref{sec:JetThicknessStudies}). Another reduction of the residual gas pressure will come with the installation of a significantly smaller vacuum chamber, which will bring apertures and pumping stages along the beam line closer to the jet. The currently installed large vacuum chamber is utilized to house multiple detector arrays around the target location to measure light reaction particles. The switch to a smaller chamber is necessary for capture reactions with SECAR to enable the installation of $\gamma$-ray detectors as close as possible to the target location.

Further developments of the JENSA system will include the installation of a gas cooler right before the nozzle. A cryogenic liquid will be cycled in a closed loop between the cooler and a refrigeration system provided by a Polycold MaxCool 4000~H Cryochiller that is already installed in the ReA3 high bay near the JENSA vacuum chamber. Gas temperatures are expected to be reduced by 100\,K, resulting in a more stable jet with less gas drifting into the surrounding vacuum chamber. In addition, JENSA is currently prepared for hydrogen operation involving the installation of additional gas analyzers, an automatic venting system, and a hydrogen safety system.

\section{Conclusions}

In summary, the JENSA gas-jet target system is currently installed in the ReA3 high bay at NSCL. An improved configuration of its components includes more powerful pumps and a reduction in the required gas volume for expensive gases to 160\,liter. JENSA provides a windowless chemically pure, highly localized gas-jet target with a high-end density of about $10^{19}$\,atoms/cm$^2$. The jet density along the gap between nozzle and receiver has been characterized in detail. Thus, a recently installed nozzle holder enables the precise determination of the vertical location of the jet that is irradiated by the incident beam. A new beam viewer at the target location provides a verification of the position of the beam.

The methods described above will be used in the future to remeasure the jet densities whenever the system configuration is changed noticeably. With a detailed knowledge of the jet characteristics, JENSA in a stand-alone mode will enable measurements of ($\alpha$,\,p), (d,\,p) neutron transfer, and ($^3$He,\,d) proton-transfer reactions in inverse kinematics with radioactive ion beams provided by ReA3. Our precise determination of the jet properties of JENSA will also enable the measurement of ($\alpha$,\,$\gamma$) and (p,\,$\gamma$) reactions with rare isotope beams provided by FRIB and ReA3 when SECAR starts operation.

\section{Acknowledgements}

Funding for the Jet Experiments in Nuclear Structure and Astrophysics gas-jet target is provided in part by the National Science Foundation under Grant Number PHY-1430152 (JINA Center for the Evolution of the Elements), Grant Number PHY-1565546 (NSCL) and Grant Number PHY-1419765 (Notre Dame). Funding was also provided by the U.S. Department of Energy (DOE), in part by the Laboratory Directed Research and Development Program of Oak Ridge National Laboratory, managed by UT-Battelle, LLC, for the U.S. DOE; by the Colorado School of Mines grants DE-FG02-10ER41704 (DOE Office of Nuclear Physics) and FG02-93ER40789 (DOE Office of Science); and by the ORNL DOE ARRA grant DE-AC05-00OR22725. The authors thank A. Snyder for his support with the assembly of the compressor's noise-isolating housing and M. Larmann for his support with the installation of the new collar and nozzle mount.


\end{document}